\documentclass[a4paper]{jpconf}
\usepackage{graphicx}
\usepackage{url}
\usepackage{iopams}

\def\araa{ARA\&A}%
\def\apj{ApJ}%
\def\apjl{ApJ}%
\def\aap{A\&A}%
\def\solphys{Sol.~Phys.}%
\def\memsai{Mem.~Soc.~Astron.~Italiana}%
\begin{document}

\title{The StaggerGrid Project: a Grid of 3D Model Atmospheres for High-Precision Spectroscopy}

\author{Remo Collet$^{1,2,3}$, Zazralt Magic$^1$, and Martin Asplund$^{1,4}$}

\address{$^1$ Max-Planck-Institut f{\"u}r Astrophysik, Karl-Schwarzschild-Str.~1, D--85741 Garching b. M{\"u}nchen, Germany}
\address{$^2$ Centre for Star and Planet Formation, Natural History Museum of Denmark
University of Copenhagen, {\O}ster Voldgade 5-7, DK--1350 Copenhagen, Denmark}
\address{$^3$ Astronomical Observatory/Niels Bohr Institute, Juliane Maries Vej 30, DK--2100 Copenhagen, Denmark}
\address{$^4$ Research School of Astronomy \& Astrophysics, Cotter Road, Weston ACT 2611, Australia}

\ead{[remo,magic,asplund]@mpa-garching.mpg.de}

\begin{abstract}
In this contribution, we present the {\sc StaggerGrid}, a collaborative project for the construction of a comprehensive grid of time-dependent, three-dimensional (3D), hydrodynamic model atmospheres of solar- and late-type stars with different effective temperatures, surface gravities, and chemical compositions.
We illustrate the main characteristics of these 3D models and their effects on the predicted strengths, wavelength-shifts, and shapes of spectral lines, highlighting the differences with respect to calculations based on classical, one-dimensional, hydrostatic models, and discuss some of their possible applications to elemental abundance analysis of stellar spectra in the context of large observational surveys.

\end{abstract}

\section{Introduction}

The Gaia mission \cite{Perryman:2001,Prusti:2011} will measure high-precision parallaxes and proper motions for $10^9$ galactic stars down to apparent magnitude $V=20$, as well as radial velocities for $10^8$ stars with $V{\leq}13$. 
In connection with the mission, a number of ground-based surveys are being planned that will carry out observations to complement the Gaia data.
Over the next five years, the upcoming Gaia-ESO public survey, for instance, will acquire high-resolution spectra for more than $10^5$ galactic stars using the VLT/FLAMES multi-fibre spectrograph.
The goal of the survey is to homogeneously derive chemical abundances for all these stars; combined with astrometry from Gaia, this information will allow to trace the most detailed and extensive chemo-dynamical map of the stellar components of the Milky Way.

Processing the enormous amount of data from this and similar surveys and extracting from them stellar elemental abundances will require fast access to grids of model stellar atmospheres and synthetic spectra covering the relevant range in terms of stellar parameters and chemical compositions.
Grids of traditional one-dimensional (1D), stationary, hydrostatic model stellar atmospheres such as {\sc MARCS} \cite{gustafsson75,gustafsson08} or {\sc ATLAS} \cite{kurucz93,kurucz05} are already available or are being re-computed together with atlases of synthetic spectra to match the specific needs of the various surveys.
The main advantage of classical model atmospheres is that the simplifying assumption of a 1D stratification allows to invest the computational resources on the solution of the radiative transfer equation for a very large number  (typically $10^5$ or more) of wavelength points. 
However, 1D models can only treat convective energy transport in an approximate manner via 
recipes such as the mixing-length theory (MLT) \cite{boehm-vitense58} or the full-spectrum-of-turbulence (FTS) model \cite{canuto91}, which are all dependent on a number of free parameters.
In late-type stars, the convective flows reach the stellar surface, affecting the atmospheric layers and, consequently, the actual spectral energy distribution in the emergent radiative flux.
Furthermore, bulk gas flows in stellar atmospheres and associated Doppler shifts also affect the broadening, shapes, wavelength shifts, and strengths of spectral lines.  
It is therefore important to properly account for their effects in order to extract accurate and precise elemental abundances from the analysis of stellar spectra.
Because traditional stationary, 1D, hydrostatic model atmospheres of late-type stars lack a consistent description of atmospheric velocity fields, Doppler broadening of spectral lines is modelled in 1D analyses by introducing additional free parameters such as micro-turbulence and macro-turbulence that generally need adjusting and tuning on an individual star basis.

In more recent years, on the other hand, a lot of efforts have been invested in the development of realistic, time-dependent, three-dimensional (3D), radiation-hydrodynamic simulations of stellar surface convection that can be directly applied as 3D model stellar atmospheres for spectral synthesis calculations (see \cite{nordlund09} for a recent review).
Three-dimensional simulations of stellar surface convection successfully reproduce important observational constraints such as the spatial properties and temporal evolution of the solar granulation pattern \cite{stein98,danilovic08,wedemeyer09}, centre-to-limb intensity variations at the surface of the Sun \cite{koesterke08,agss09} as well as other nearby stars \cite{aufdenberg05,bigot06}, as well as the detailed shapes and wavelength shifts of spectral lines in solar- and late-type stars \cite{nordlund90,allende02a,Ramirez:2010}.
At present, a number of codes suitable for constructing realistic 3D hydrodynamic model atmospheres are available and actively developed. 
A few prominent examples in this respect are the {\sc Stagger} \cite{nordlund95}, {\sc Bifrost} \cite{Gudiksen:2011}, {\sc CO$^5$BOLD} \cite{freytag02}, and {\sc MURaM} \cite{Vogler:2005} codes.
Three-dimensional models have recently started to be employed for spectroscopic stellar abundance analyses \cite{asplund99,asplund03,collet06,caffau10,gonzalez10}. 
Similarly as in the case of 1D models, organized grids of 3D model atmospheres are currently being developed and becoming available for this purpose. 
The recently presented {\sc CIFIST} collection of {\sc CO$^5$BOLD} model atmospheres \cite{ludwig09}, for instance, is a pioneering example of such a grid.
In this contribution, we will present the general outlines of the {\sc StaggerGrid}, an alternative project for the construction of a grid of 3D model stellar atmospheres of late-type stars with the {\sc Stagger-Code}, and we will discuss some of its possible applications in the context of spectral line formation and elemental abundance analysis.

\section{The StaggerGrid}

\begin{figure}
\centering
\resizebox{0.60\hsize}{!}{ \includegraphics{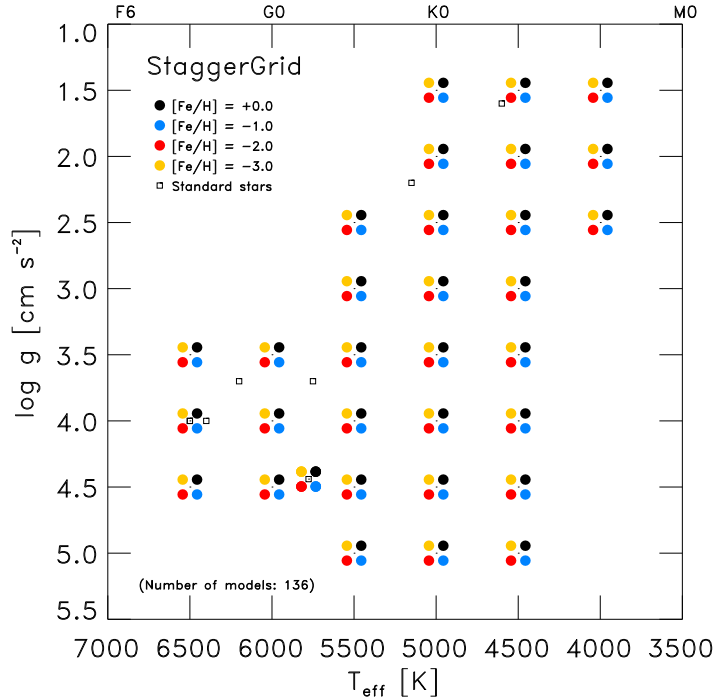}  }
\caption{Overview of the {\sc StaggerGrid}: the dots indicate the effective temperatures ($T_\mathrm{eff}$) and surface gravities ($\log{g}$) of the 3D surface convection simulations being computed for the project. The coloured circles represent the different compositions (scaled solar metallicity \cite{agss09} with [Fe/H] ranging from $+0$ to $-3$, with a $\alpha$-enhancement value of [$\alpha$/Fe]$=+0.4$ for the metal-poor models).  
The empty squares indicate the $T_\mathrm{eff}$- and $\log{g}$-values adopted for the surface convection simulations of the Sun and other standard stars important for stellar spectroscopy (see also Table~\ref{tab:standard}).}
\label{fig:staggergrid}
\end{figure}

We are using a custom version of the MPI-parallel {\sc Stagger-Code} to carry out time-dependent, three-dimensional, radiation-hydrodynamic simulations of convection at the surface of late-type stars for a range of effective temperatures, surface gravities, and metallicities.
The general outline of the grid in stellar parameter space is shown in Fig.~\ref{fig:staggergrid}.

The simulations are of the so-called \emph{star-in-the-box} kind: each simulation's domain is a rectangular, three-dimensional volume located at the surface of the star, with periodic boundaries horizontally and open boundaries vertically.
The mass, momentum, and energy conservation equations for a compressible, viscous flow are discretized and solved on a Cartesian numerical mesh with $240{\times}240{\times}240$ grid-points as default numerical resolution.
The code uses artificial viscosity operators to deal with the effects of numerical diffusion. The parameters for these operators have been tuned using solar surface convection simulations to minimize such effects without causing undesired or excessive smoothing of shocks. Once tuned, these parameters are kept fixed for all models. Changing these parameters does affect the sharpness of, e.g., temperature and density inhomogeneities but essentially does not alter the main features of the simulations, such as the basic morphology of the convective flows or the average temperature and density stratifications.

We choose the physical size of the domains so that the simulations typically host about ten convective granules at any given time and cover about twelve pressure scale heights vertically, extending from ${\log}\,{\tau}_\mathrm{Ross}{\leq}{-4}$ to ${\log}\,{\tau}_\mathrm{Ross}{\geq}{+6}$ in terms of Rosseland optical depth.
In the setup currently used for the {\sc StaggerGrid} models, we assume a constant vertical gravitational acceleration in the simulation box and neglect sphericity effects.
The physical size of the computational domain of low-surface-gravity {\sc StaggerGrid} models is, however, still sufficiently small compared with the stellar radius, implying that these approximations are acceptable.
At the bottom boundary, located deep below the surface, in the convectively unstable layers, we impose constant pressure across the whole layer and require the inflowing gas to have constant entropy per unit mass.

Our goal is to construct model atmospheres that can be used for accurate abundance determinations, so we try to include as realistic input physics as possible in the simulations: we implement a state-of-the-art equation-of-state \cite{mihalas88} and up-to-date continuous  \cite{gustafsson75} (see also Trampedach et al., in prep.) and  line opacities \cite{gustafsson08}.  

In order to model the temperature stratification in the outer stellar layers properly, it is paramount to account for the energy exchange between gas and radiation.
At each time-step during the simulation, we solve the radiative transfer equation along the vertical as well as  eight other inclined directions (two $\theta$- and four $\phi$-angles) using a Feautrier-like method \cite{Feautrier:1964} and compute the necessary radiative heating rates for the energy conservation equation.
The choice of a Feautrier-like method implies that each direction is effectively counted twice, once for outgoing rays and once for incoming rays.

Temperature gradients at the surface of red giant models can become very steep. 
In general, a fixed geometrical depth scale does not allow to properly resolve these gradients, which can give rise to numerical artifacts in the outgoing intensity pattern at the surface. 
In our version of the {\sc Stagger-Code}, we have therefore implemented an adaptive depth scale to solve this issue (Collet et al., in prep.).

In order to reduce the computational burden for the solution of the radiative transfer equation, we approximate the source function with the Planck function at the local temperature ($S_\nu=B_\nu(T)$) and neglect the contribution of scattering to the total extinction in the optically thin layers \cite{collet11}.
Furthermore, we use a multi-group or opacity-binning approximation \cite{nordlund82,skartlien00} to account for the dependence of opacity on wavelength: we sort the monochromatic opacities into groups (or \emph{bins}) according to wavelength range and opacity strength, then solve the radiative transfer equation for the individual group mean opacities and the integrated group source functions \cite{collet11}.
We calibrate the opacity-binning method for each specific choice of stellar parameters in order to achieve an as accurate as possible representation of the heating rates. 
We normally do that by finely adjusting the criteria for opacity-bin-membership for each individual simulation until the difference between the heating rates computed with opacity-binning and with the full monochromatic solution for the average stratification from the 3D model is minimized.

\begin{center}
\begin{table}[h]
\centering
\caption{\label{tab:standard} Stellar parameters of some standard stars for which 3D hydrodynamic {\sc Stagger-Code} models have been computed as part of the grid. For all stars, we have assumed a standard solar composition with the abundances of metals scaled proportionally to the relevant [Fe/H], with $\alpha$-enhancement of [$\alpha$/Fe]$=+0.4$ for the metal-poor models. In addition, for HE1327$-$2326 and HE0107$-$5240, we have also accounted for the peculiar CNO-enhancement of these stars.} 

\begin{tabular}{lcccc} 
\br
Star & $T_\mathrm{eff}/[\mathrm{K}]$& $\log{g/[\mathrm{cm s}^{-2}]}$ & [Fe/H] \\
\mr
Sun			& $5780$ & $4.44$ & $+0.0$ \\
Procyon		& $6500$ & $4.00$ & $+0.0$ \\
HD140283		& $5750$ & $3.70$ & $-2.5$ \\
HD84937		& $6400$ & $4.00$ & $-2.0$ \\
G64-12		& $6500$ & $4.00$ & $-3.0$ \\
HD122563		& $4600$ & $1.60$ & $-3.0$ \\
HE1327$-$2326	& $6200$ & $4.00$ & $-5.0$ \\
HE0107$-$5240	& $5200$ & $2.20$ & $-5.0$ \\
\br
\end{tabular}
\end{table}
\end{center}

At the present time, a grid of models using a radiative transfer solution with six opacity bins is nearing completion. 
In addition, we have also computed models for some reference stars (see Table~\ref{tab:standard}) using a more refined opacity-binning scheme with twelve opacity bins. 
We plan to eventually extend the twelve-bin opacity binning scheme to the computation of all {\sc StaggerGrid} models.

To produce the initial 3D snapshot of a simulation for a given set of targeted stellar parameters, we take the physical structure from another simulation snapshot previously computed for other stellar parameters and scale it appropriately. We do that by looking at the ratios of spatial scales and of various important physical quantities such as temperature, density, and pressure from 1D model envelopes corresponding to the same stellar parameters.
The scaled model is then allowed to adjust and relax. 
Spurious p-mode-like oscillations are then damped and filtered out to rid the simulation from excess energy caused by the imperfections of the scaling procedure and to allow only the natural modes of oscillation to survive.

\section{Results and applications}

\begin{figure}
\centering
\resizebox{0.60\hsize}{!}{ \includegraphics{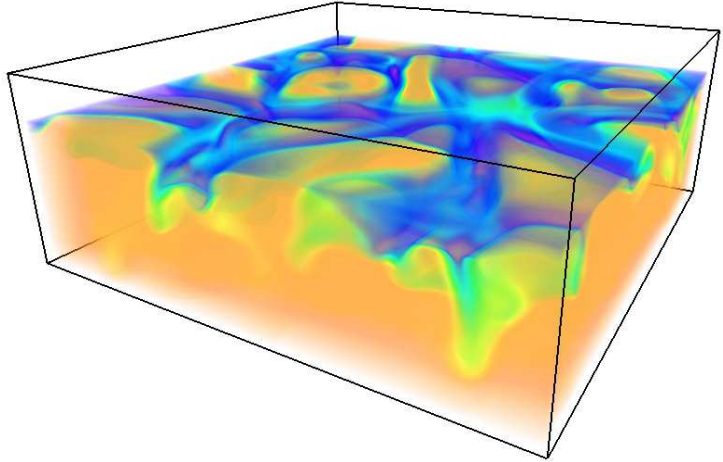}  }
\caption{Specific entropy per unit mass for a metal-poor red giant simulation: red-orange hues indicate high-entropy regions, purple-blue-green hues low-entropy ones. The physical size of the box is about $1200{\times}1200$~Mm${^2}$ horizontally and $450$~Mm vertically. The warm inflowing gas from the bottom of the simulation domain has constant, high, specific entropy per unit mass. As the ascending gas reaches the optically thin layers at the surface, it rapidly cools via radiation losses, lowering its entropy. It eventually becomes denser than the surrounding material and falls back toward the interior of the star in narrow downdrafts.}
\label{fig:entropy}
\end{figure}

As an illustration of the typical physical structures resulting from 3D simulations, in Fig.~\ref{fig:entropy}, we show the gas entropy per unit mass in a representative snapshot of a {\sc Stagger-Code} red-giant surface convection simulation.
The apparent characteristic surface convective pattern with large granules with warm, ascending gas
surrounded by an intergranular network of cooler, denser, downflowing material 
emerges naturally from the simulations without the need for adjustable parameters.
The first fundamental difference between 3D and 1D model stellar atmospheres is therefore that the 3D models self-consistently predict the emergence of density and temperature inhomogeneities at the stellar surface and their correlation to macroscopic velocity fields.
The non-linear dependence of the populations of energy levels of atoms and molecules as well as of ionization balance and molecular equilibrium on such inhomogeneities ultimately leads to appreciable differences between the profiles and strengths of spectral lines generated using 3D and 1D models, even in those cases where the 1D and the average 3D stratifications are not too dissimilar from each other \cite{collet07}.

Another difference, especially important in the context of spectral line formation calculations, is that 1D model atmospheres of late-type stars generally predict a steeper temperature stratification as a function of optical depth than 3D simulations at the optical surface and in the layers immediately below it. 
The effects of this are apparent, for instance, in the different predictions of centre-to-limb variations (CLV) in the outgoing continuum radiation intensity. In the Sun's case, in particular, the predicted CLV can be tested directly against observations, showing that the 3D simulations, contrary to 1D models, can successfully reproduce such variations across the solar spectrum \cite{agss09}.

\begin{figure}
\centering
\resizebox{\hsize}{!}{ 
\includegraphics{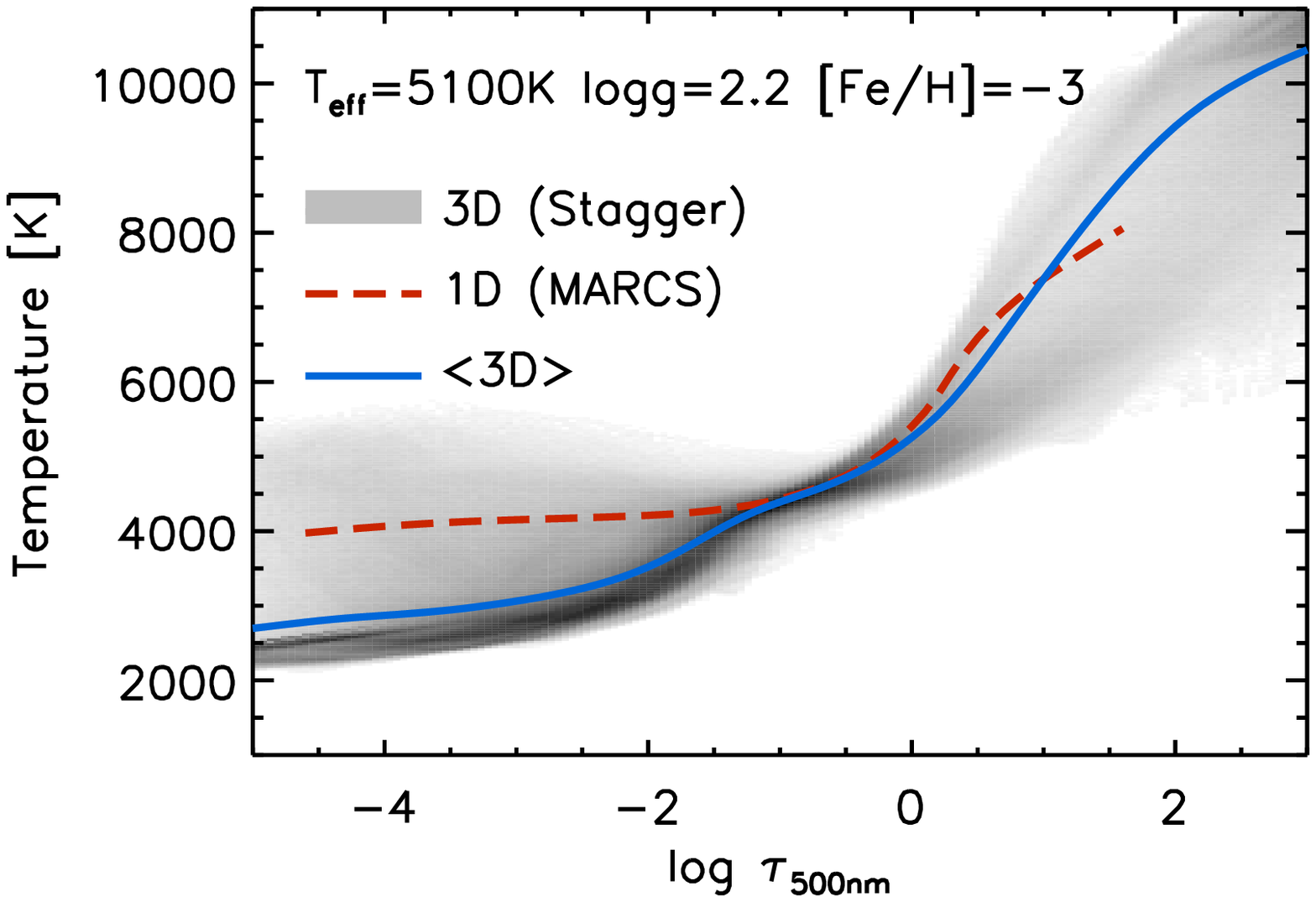}  
\includegraphics{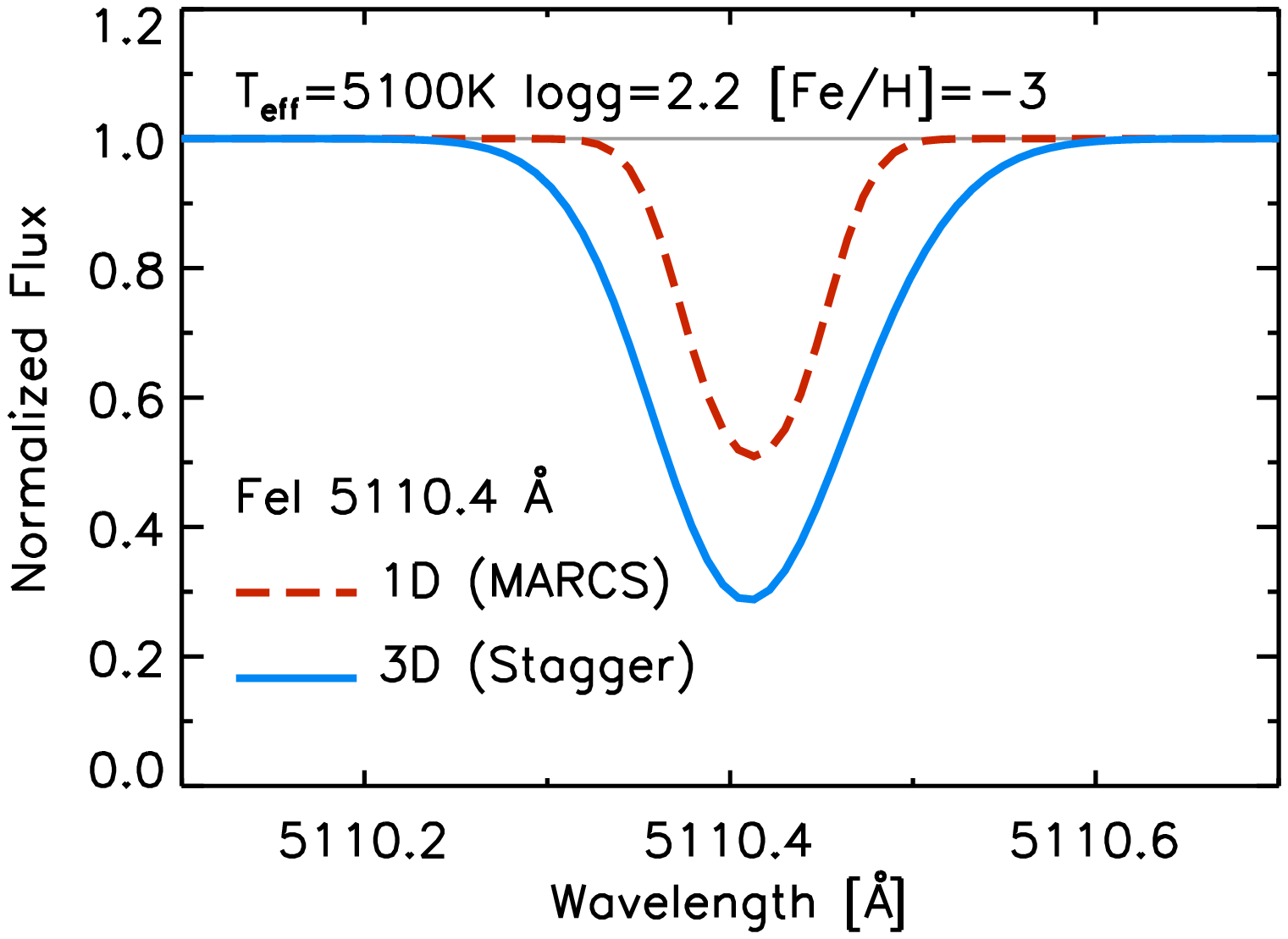}  
}
\caption{\emph{Left panel, grey shaded area:} Temperature distribution as a function of optical depth in a representative snapshot from a 3D {\sc Stagger-Code} simulation of metal-poor red giant; \emph{blue line:} mean temperature stratification; \emph{red dashed line:} temperature stratification from corresponding 1D {\sc marcs} model. \emph{Right:} synthetic profiles for the Fe~{\sc i} (neutral iron) spectral line at $5110.4$~{\AA} computed in local thermodynamic equilibrium (LTE) with the 3D (\emph{blue line}) and 1D (\emph{red dashed line}) models of red giant stellar atmospheres shown in the left panel, assuming the same Fe abundance.}
\label{fig:temp-profile}
\end{figure}

Finally, a third important difference is that 3D stellar surface convection simulations of metal-poor late-type stars predict significantly cooler upper-atmospheric temperature stratifications than 1D models \cite{asplund99,collet07} (Fig.~\ref{fig:temp-profile}, left panel).
In the 3D hydrodynamic case, the temperature in these layers is essentially regulated by two mechanisms: radiative heating due to reabsorption of continuum-radiation by spectral lines and adiabatic cooling associated with diverging flows above granules.
At low metallicities, the contribution of spectral lines to the total opacity in the upper atmosphere is relatively weak: the significance of radiative heating in these layers is therefore reduced relative to adiabatic cooling and the temperature balance is shifted to lower temperatures.
Stationary, 1D, hydrostatic model atmospheres do not account for such adiabatic cooling component associated with gas expansion. The thermal balance in such models is regulated purely via heating and cooling by radiation, ultimately resulting in artificially high temperatures compared with the 3D case.

Differences between the temperature stratifications in the upper atmosphere of 1D and 3D models can amount to ${\sim}1000$~K in some cases, severely affecting the excitation, ionization, and molecular equilibria in those layers.
Such temperature differences can have a particularly large impact on the strengths of synthetic line profiles from temperature-sensitive species and, consequently, on the elemental abundances that can be derived by comparing theoretical and observed profiles.
As an example, Fig.~\ref{fig:temp-profile}, right panel, shows the predicted profile of a neutral Fe line, computed in local thermodynamic equilibrium (LTE) using a 3D model atmosphere of a very metal-poor red giant star and its corresponding 1D counterpart, assuming the same Fe abundance in the two cases. 
Because of the line's temperature-sensitivity,  the resulting 3D line profile is significantly stronger than 
the 1D one.
This also implies that a 3D LTE analysis of this line would require in this case a significantly lower Fe abundance than a 1D LTE analysis to reproduce the strength of a given, observed, line profile.
Differences between the derived 3D and 1D abundances can be of the order of $-0.5$~dex or even larger in magnitude for lines from neutral atoms and molecules \cite{collet07,collet09}.
Based on the experience with previous simulations carried out by our group \cite{asplund01,collet07} with a predecessor of the {\sc Stagger-Code} \cite{stein98} and from preliminary analyses with the present {\sc StaggerGrid} models, we see that the 3D$-$1D LTE abundance differences derived from lines from molecules or neutral atoms typically tend to increase in magnitude (i.e., become more negative) when the metallicity decreases, or when the effective temperature increases, or, also, when the surface gravity decreases.
At the moment, however, these results are still based on a limited number of tests restricted to some parts of the grid. We are therefore planning to carry out a more systematic comparison of the results of abundances determinations with 3D and 1D models and study in greater detail the trends of the 3D$-$1D abundance differences with stellar parameters.

\section{Comparison with other models}
The {\sc Stagger} and {\sc CIFIST} grids are conceptually similar in terms of basic structure and purposes. Both grids also rely on essentially the same opacity sources and use up-to-date, realistic equation-of-state packages and input physics.
The main differences are in the adopted numerical methods (codes), basic resolution, and physical extension of the models (current box-in-the-star {\sc StaggerGrid} models use a higher numerical resolution and generally extend down to deeper layers) and in the implementations of radiative transfer and opacity binning. The latter, in particular, may be responsible for the apparent differences between the resulting temperature stratifications from simulations of metal-poor stars. The {\sc StaggerGrid} metal-poor simulations predict a cooler temperature stratification in the upper-photospheric layers compared with analogous {\sc CIFIST} models computed for the same stellar parameters \cite{bonifacio09,collet11,Ludwig:2011}. A systematic comparison of the two grids has not been carried out yet, but it is being planned.
However, it is important to mention that, in spite of the differences between the two grids at low metallicity, the {\sc StaggerGrid} and {\sc CIFIST} solar surface convection simulations are actually in very good agreement with each other, as well as with the current solar simulation by the {\sc MURaM} group \cite{Beeck:2011}.

\section{Summary and outlook}
In light of the results we have presented here, accounting for the differences between 3D and 1D models is paramount in order to  accurately determine elemental abundances and other stellar properties from the analysis of stellar spectra.
Once the grid will be completed, we will therefore compute synthetic spectra for all 3D models and their 1D counterparts, covering the range from ultraviolet to infrared wavelengths.
These spectra will be used for computing synthetic colours, deriving stellar parameters and abundances, studying the properties of stellar surface convection across the H-R diagram, and for many other applications (see, e.g., Chiavassa et al., these proceedings).

We will also construct and make publicly available average 3D stratifications from the {\sc StaggerGrid} models. The information from the full 3D structures will also be used to provide physical constraints to free parameters used in 1D analyses such as micro- and macro-turbulence.
This will facilitate the implementation of the main results from 3D modelling in existing, commonly used, 1D spectral line formation packages.
This would be particularly useful in order to systematically and consistently study the combined effects of granulation and departures from local thermodynamic equilibrium in late-type stellar atmospheres (see Bergemann et al., these proceedings).

In conclusion, the {\sc StaggerGrid} will offer a powerful and flexible tool for progressing toward precise and accurate analyses of stellar spectra and elemental abundances determinations, and, when combined with Gaia-related follow-up surveys, it will provide a significant leap forward in our understanding of Galactic chemical evolution.

\section{Acknowledgments}
The {\sc StaggerGrid} project is a collaboration involving scientists from several institutes including the Max Planck Institute for Astrophysics (MPA, Garching), the Niels Bohr Institute (NBI, Copenhagen), the Observatoire de la C{\^o}te d'Azur (OCA) in Nice, the Institute of Astronomy and Astrophysics (IAA) in Brussels, and the Joint Institute for Laboratory Astrophysics (JILA, Boulder).

\section*{References}
\bibliographystyle{iopart-num}
\providecommand{\newblock}{}

\end{document}